\documentclass{mpe_report}

\usepackage{psfig,graphicx,epsfig}
\usepackage{color}
\usepackage{amsmath,amssymb,epic,eepic,array}

\begin{document}

\pagenumbering{arabic}
\setcounter{page}{157}

 \renewcommand{\FirstPageOfPaper }{157}\renewcommand{\LastPageOfPaper }{160}
\def\Bs{B_{\rm s}}
\def\Bd{B_{\rm d}}
\def\Ts{T_{\rm s}}
\def\be{\begin{equation}}
\def\ee{\end{equation}}
\def\lesssim{\raisebox{-0.3ex}{\mbox{$\,\, \stackrel{<}{_\sim} \,\,$}}}
\def\gtrsim{\raisebox{-0.3ex}{\mbox{$\stackrel{>}{_\sim} \,$}}}
\def\EB{\hbox{${\rm {\bf E} \times {\bf B}}$}}

\title{A Model for unusual pulsed X-ray emission from \\ PSR J1119 - 6127}
\author{George I. Melikidze \inst{1,2}, Janusz Gil\inst{1}\and Andrzej Szary\inst{1}}
\institute{Institute of Astronomy, University of Zielona G\'ora, Lubuska 2, 65-265, Zielona G\'ora, Poland \and
Abastumani Astrophysical Observatory, Al. Kazbegi ave. 2a, 0160, Tbilisi, Georgia} 

\authorrunning{Melikidze et al.}
\titlerunning{A Model for unusual pulsed X-ray emission from}
\maketitle

\begin{abstract}
We propose a model, which naturally explains an unusual thermal X-ray emission, observed from PSR J1119-6127. The model
is based on the assumption that the pulsar magnetic field near the stellar surface differs significantly from the pure
dipole field. We show that the structure and curvature of the field lines can be of the kind that allows the pair
creation in the closed field line region. The created pairs propagate along the closed field lines and heat the stellar
surface beyond the local poles. It is demonstrated that such a configuration can be easily realized.
\end{abstract}


Thermal X-ray emission seems to be a quite common feature of the radio pulsars. On the other hand characteristics of
such radiation allows us to get a lot of information about the polar cap region of the pulsars. Standard model of the
radio pulsars assumes that there exists the Inner Acceleration Region (IAR) above the polar cap where the electric
field has a component along the magnetic field lines. The particles (electrons and positrons) are accelerated in both
directions: outward and toward the stellar surface. Consequently, outstreaming particles generate the magnetospheric
(radio and high-frequency) emission while the the backstreaming particles heat the surface and provide necessary energy
for the thermal emission. In such a scenario X-ray diagnostics seems to be an excellent method to get insight into the
most intriguing regain of the neutron star.

The black body fit allows us to obtain directly the bolometric size ($A_{\rm bol}$) and temperature ($T_{\rm s}$) of
the polar cap. In most cases $A_{\rm bol}$ is much less than the conventional polar cap area. It can be easily
explained by assuming that the surface magnetic field of pulsars differs significantly from the pure dipole one. Then,
one can estimate an actual surface magnetic field by the magnetic flux conservation law as $b=A_{\rm bol}/A_{\rm
pc}=B_{\rm d}/B_{\rm s}$. Here $B_{\rm d}=2\times10^{12}\left( P {\dot P}_{15} \right)^{1/2}$, $P$ is the pulsar period
in seconds and ${\dot P}_{15}={\dot P}/10^{15}$ is the period derivative. In most cases $b\sim 10 - 60$, which implies
$B_{\rm s}>>B_{\rm d}$, while $T_{\rm s}\sim (2 - 4)\times 10^6 $ K. Recently Gil, Melikidze \& Zhang (2005, 2006,
GMZ06 henceforth) have shown that the model of the Partially Screened Gap (PSG) can interpret the observational data
not only qualitatively but also quantitatively.

The X-ray observations of PSR J1119-6127 showed quite unusual features of this pulsar. As it was demonstrated by
Gonzalez et al. (2005, G05 hereafter) the {\it XMM-Newton} observations denote the thermal feature of the pulsed X-ray
emission from this pulsar. The derived characteristics of the black body fit are as follows: $A_{\rm
bol}=3.6^{+4.9}_{-0.6}\times 10^{13}$ cm$^2$ and $T_{\rm s}=2.4^{0.3}_{-0.2}\times 10^6$ K. The X-ray flux is estimated
as $L_{\rm x}=2.0^{+2.5}_{-0.4}\times10^{33}$ erg s$^{-1}$. Let us note that both $A_{\rm bol}$ and $L_{\rm x}$ depend
on the distance estimation, which for this pulsar is estimated as $D=8.4\pm 0.4$ kpc (Caswell et al. 2004), while
Cordes-Lazio NE2001 (2002) Electron Density Model suggests the distance estimate as $D=17$ kpc, $(10 < D < 50 )$.
Therefore, if the distance is underestimated the flux as well as $A_{\rm bol}$ are even larger.

PSR J1119-6127 was discovered in the Parkes Multibeam Pulsar Survey (Camilo et al. 2000). The observed parameters of
the pulsar are as follows: $P=0.408$ s, $P {\dot P}_{15}=4.0\times10^3$, $B_{\rm d}=8.2\times10{13}$ (note a slight
inaccuracy in estimation of $B_{\rm d}$ by G05) and the spin-down energy loss $L_{\rm sd}=2.3\times 10^{35}$ erg
s$^{-1}$. The conventional polar cap area is about $A_{\rm pc}=1.6\times10^9$ cm$^2$. As we see the efficiency of X-ray
emission defined as $\xi=L_{\rm x}/L_{\rm sd}\sim 0.009$ is of the same order of magnitude as it is for other pulsars,
while the bolometric area $A_{\rm bol}$ exceeds $A_{\rm pc}$ $2\times 10^3$ times.

\section{The model}

\begin{figure}
\centerline{\psfig{file=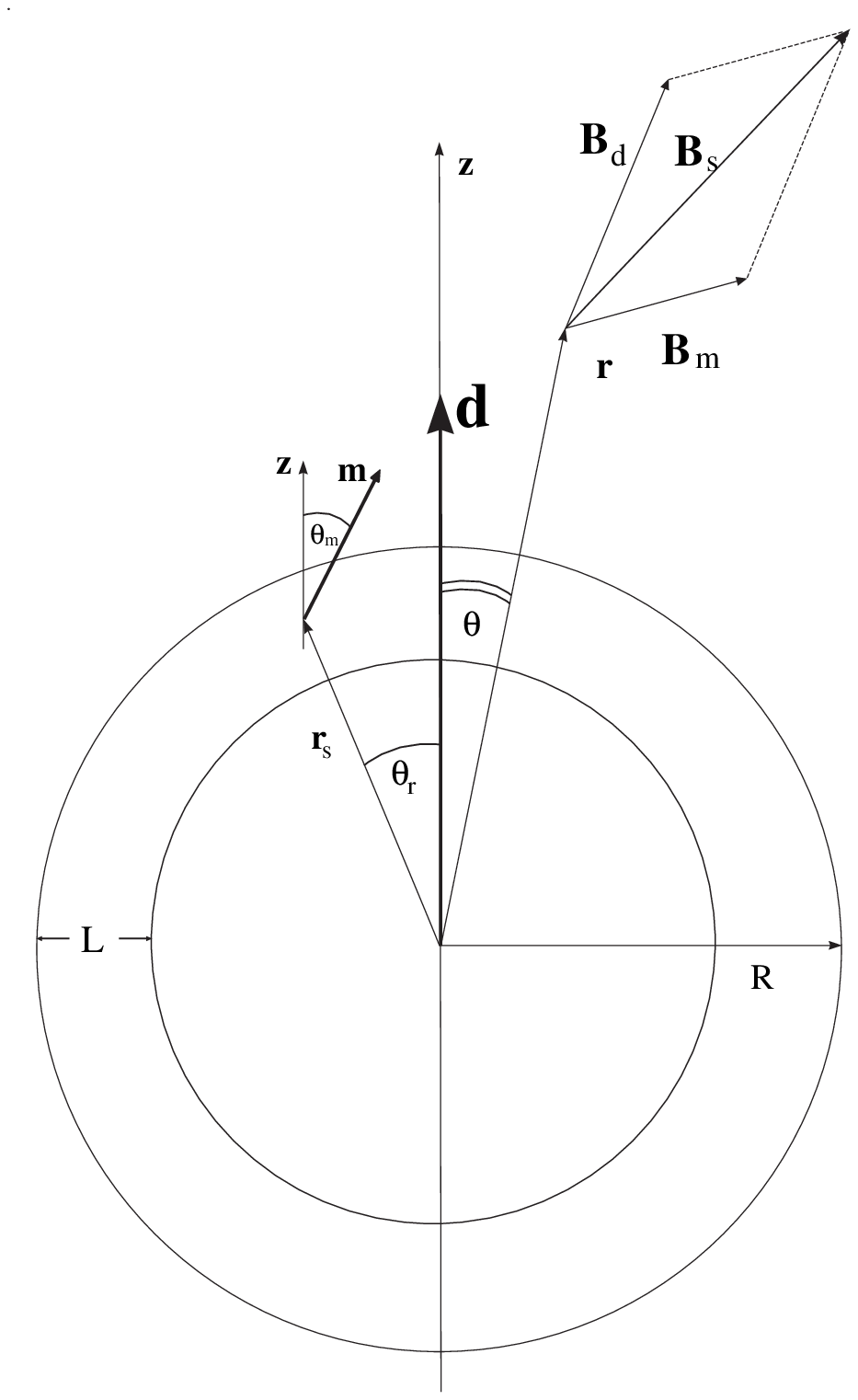,width=8.8cm,clip=} }
\caption{Superposition of the star centered global magnetic dipole ${\bf d}$ and crust anchored local dipole ${\bf m}$
placed at ${\bf r}_s = (r_s \sim R; \theta = \theta_r )$ and inclined to the z-axis by an angle $\theta_m$. The actual
surface magnetic field at radius vector ${\bf r} = (r; \theta)$ is ${\bf B}_s = {\bf B}_d+{\bf B}_m$, where ${\bf B}_d
= 2{\bf d}/r^3$, ${\bf B}_m = 2{\bf m}/|{\bf r}-{\bf r}_s|$, $r$ is the radius (altitude) and $\theta$ is the polar
angle (magnetic colatitude). $R$ is the radius of the neutron star and $L$ is the crust thickness.
\label{model}}
\end{figure}

We propose the model, which is based on the assumption that the surface magnetic field of pulsar differs greatly from
the dipolar field. The curvature and strength of the surface field can be much bigger than those which follow from the
simple dipolar geometry. This assumption seems to be quite natural and as it was mentioned above supported by the X-ray
observations of the rotation-powered pulsars (GMZ06). Gil, Melikidze, Mitra (2002) proposed that the actual surface
magnetic field can be modelled as a superposition of the global dipole field and crust-anchored small scale magnetic
anomalies. They also provided corresponding numerical formalism. Such an approach allows to get some insight on the
geometry, as well as to estimate the curvature of the field lines. Fig.~\ref{model} represents the general picture of
the model.

\begin{figure}
\centerline{\psfig{file=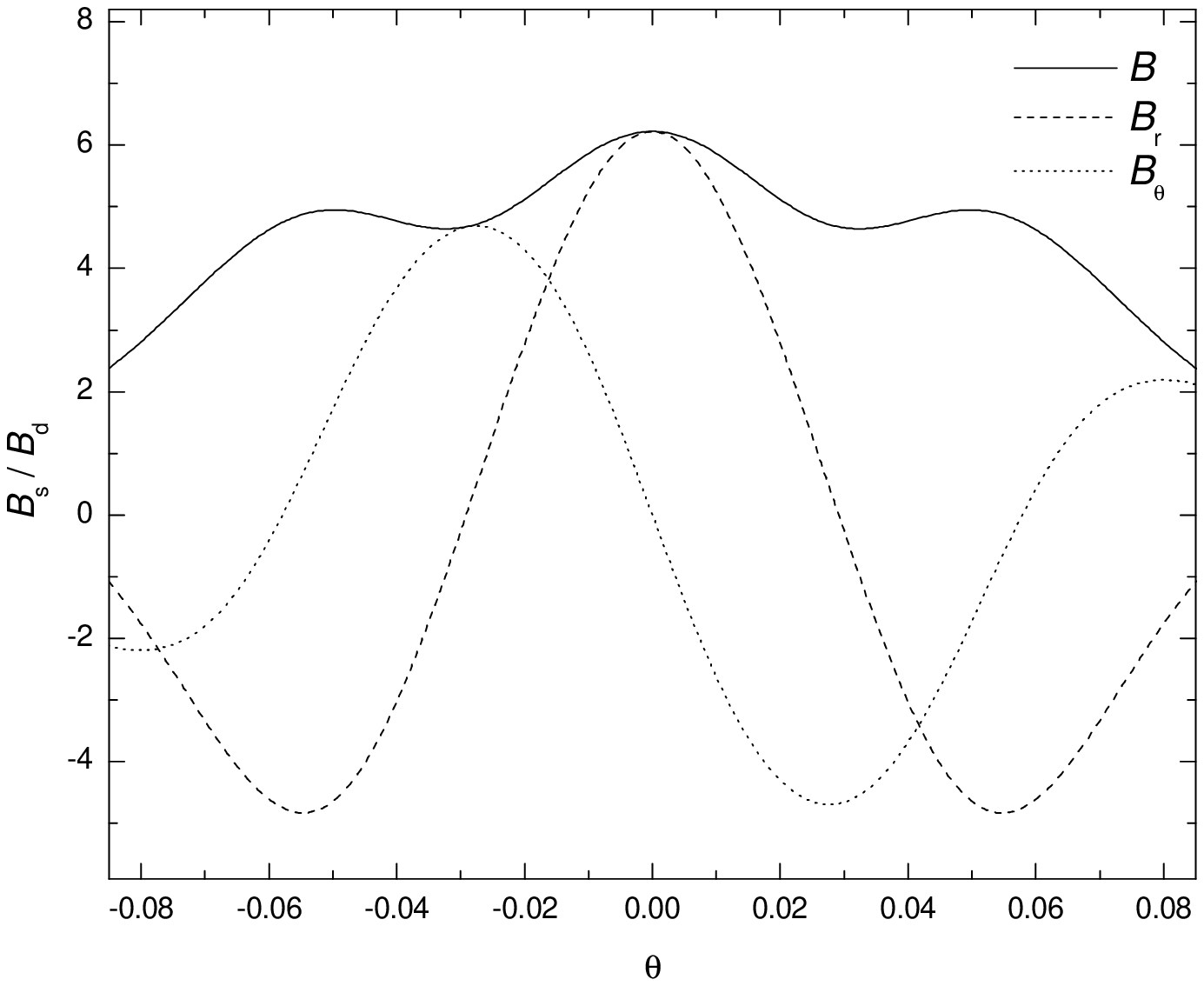,width=8.8cm,clip=} }
\caption{The magnetic field and its components near the stellar surface at the polar cap region.
\label{magfieldsurface}}
\end{figure}

\begin{figure*}
\centerline{\psfig{file=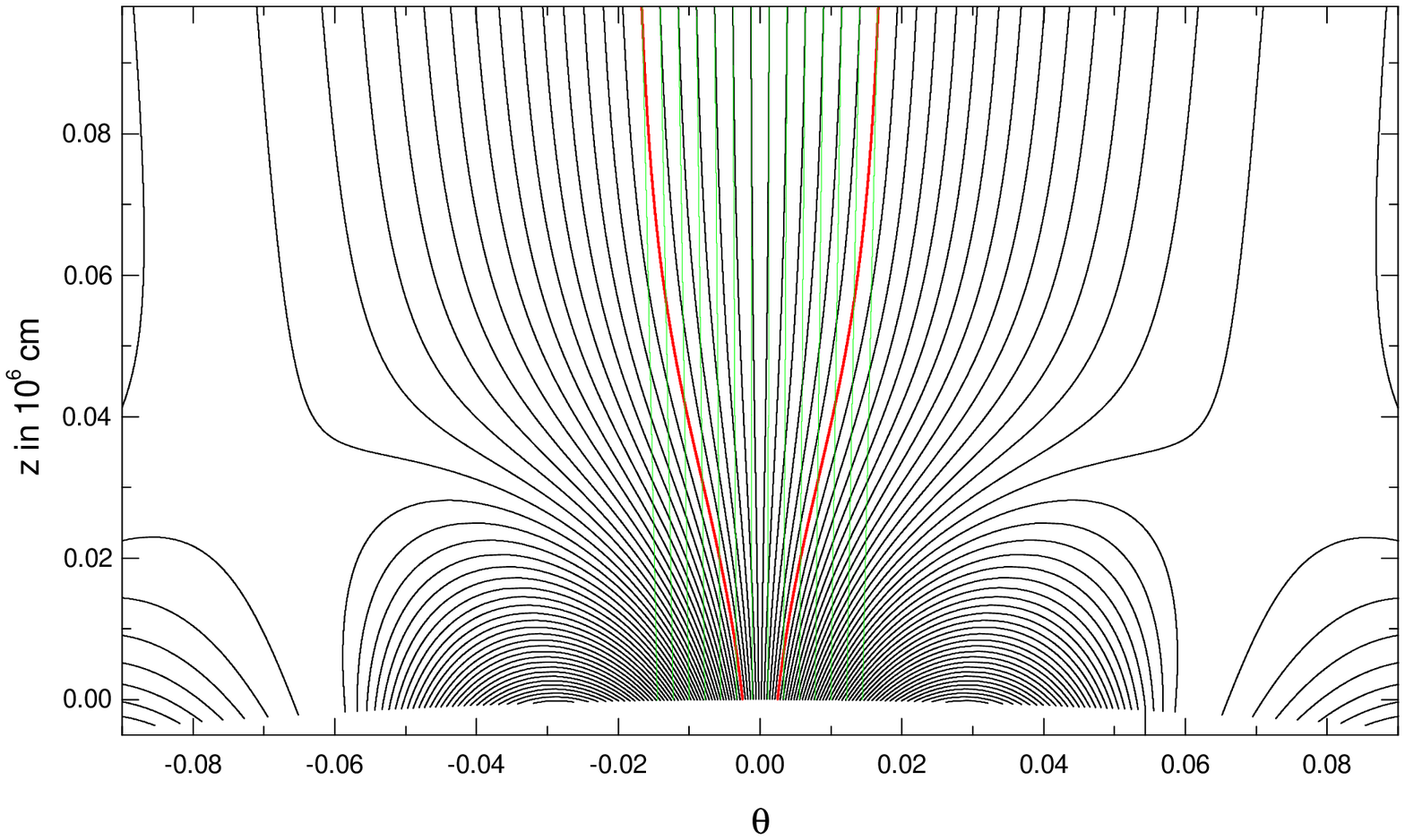,width=16cm,clip=} }
\caption{Cartoon of the magnetic field lines in the polar cap region. There are three crust anchored magnetic
anomalies in this case: the central one is aligned with the global dipole, while two others are directed to the
opposite direction. Distance between the local dipoles is 500 meters. The green lines represent the pure dipole filed.
The red lines correspond to the last open field lines, which at high altitudes coincide with the dipole field lines.
$\theta$ is the magnetic colatitude in radians.
\label{cartoon}}
\end{figure*}

\begin{figure}
\centerline{\psfig{file=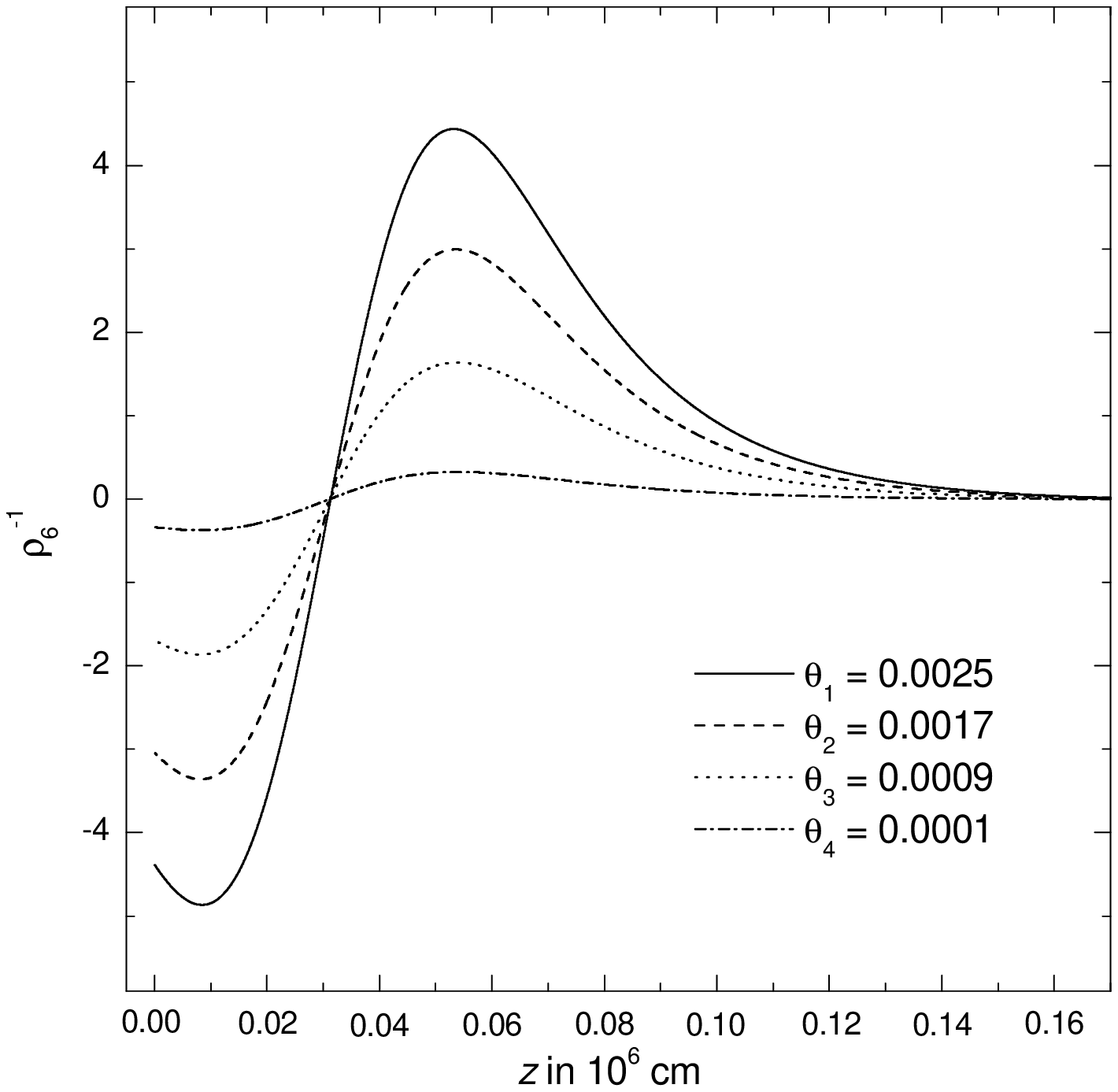,width=8.8cm,clip=} }
\caption{Curvature ($1/\rho_6$) of the field lines vs of the altitude.
Various $\theta$-s define the footpoints of the corresponding field lines at the stellar surface.
\label{curvature}}
\end{figure}

Fig.~\ref{cartoon} is a cartoon of the magnetic field lines for any general symmetric case (see figure caption for
details). The corresponding magnetic field strength is shown in Fig.~\ref{magfieldsurface}, while Fig.~\ref{curvature}
shows the curvature of the open magnetic field lines as a function of the altitude. As it is clear from
Fig.~\ref{curvature} the crust-anchored magnetic anomalies can alter not only a value of the curvature but also its
sign. In the pure dipole configuration photos radiated almost tangent to the magnetic field lines always propagate
inward, i.e. in the direction of the global dipole axes. Therefore, pair creation can occur only in the open field line
region. However, in the presence of the local anomalies the field configuration changes drastically. Despite the fact
that Fig.~\ref{model} represents some particular case, the change of the curvature sign is quite general.

\begin{figure}
\centerline{\psfig{file=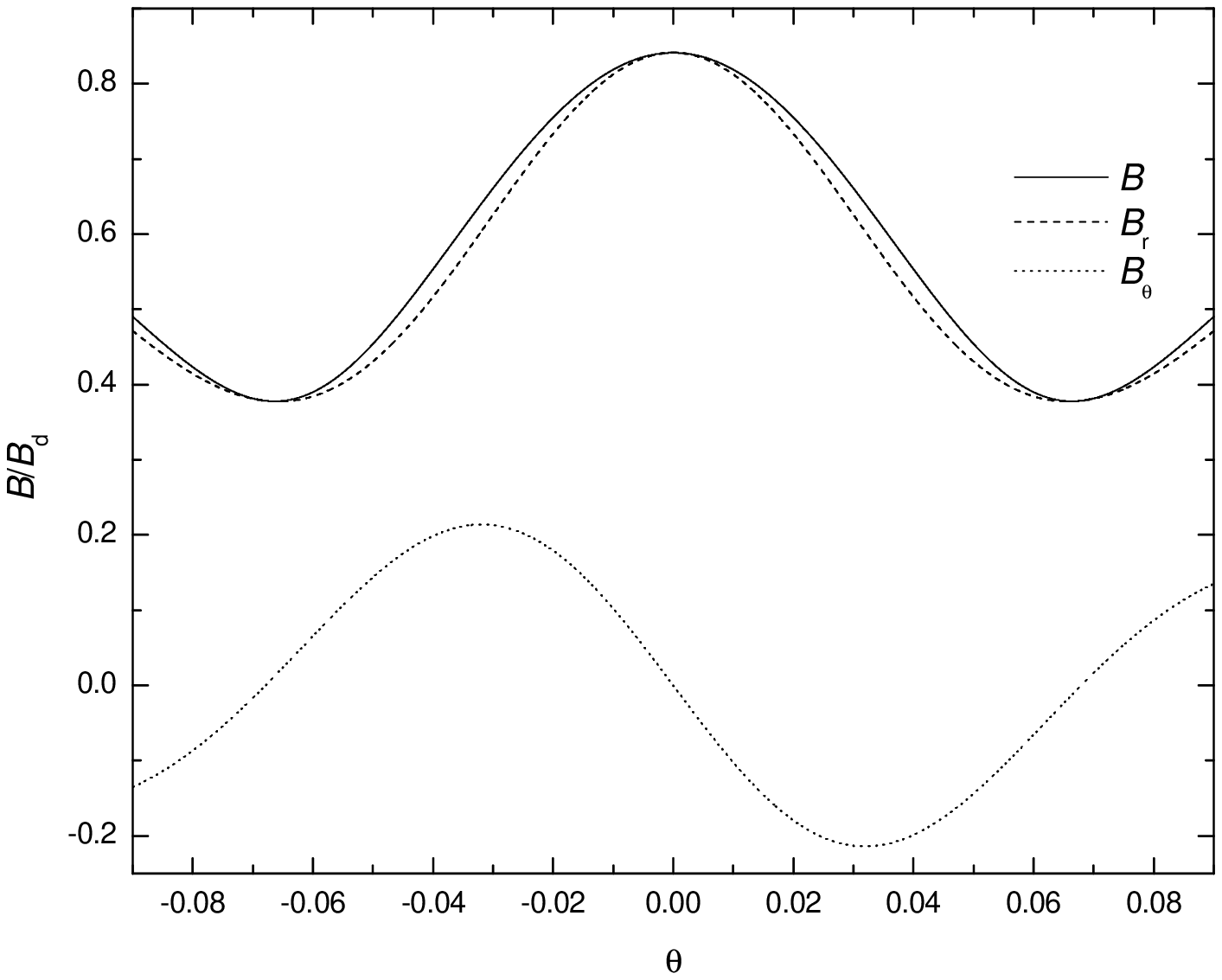,width=8.8cm,clip=} }
\caption{The magnetic field and its components at the altitude that corresponds to the strongest
curvature of the field lines.
\label{magfieldup}}
\end{figure}

\section{Estimation of the particles Lorentz factor}

Let us estimate the characteristic Lorentz factors of the particles accelerated in IAR. Following GMZ06 we use PSG
model for IAR. The efficiency of the thermal X-ray emission from the drifting pulsars is quite well explained by this
model (GMZ06). In addition GMZ06 gives a proper estimate for the surface temperature if the bolometric surface
coincides with the actual polar cap surface. In the case under consideration the latter condition is not satisfied.
Therefore, we should estimate the screening factor (see GMZ06 for details) using only efficiency of the X-ray emission,
while the temperature and $A_{\rm bol}$ should be defined by the heating-cooling balance.

The mean Lorentz factor of the accelerated particles is the main parameter that defines the energy outflow from the
polar cap. The Near Threshold (NT) condition (Gil \& Melikidze 2002) implies that the energy of the pair-creating
photon should be
\be
\hbar\omega=2m_e c^2/sin\vartheta\sim 1.6\times10^{-3}\rho_6/h_3.
\label{eq1}
\ee
Here $h_3\equiv h/(10^3 {\rm cm})$, $h$ is the gap height and $\rho_6\equiv \rho/(10^6 {\rm cm})$ $\rho$ is the
curvature radius of the local magnetic field lines. In the gap two kinds of photons can generally exist: the photons
(with energy $\hbar \omega_{\rm cr}\sim1.4\times\gamma^3/\rho_6$ ) emitted by the curvature radiation and/or the
photons (with energy $\hbar\omega_{\rm ics}\sim \gamma b$ ) generated by the inverse Compton scattering process. For
the simplicity we only consider the curvature radiation dominated PSG. The energy of the curvature photons writes:
\be
\hbar\omega_{\rm cr}\sim1.4\times10^{-5}\gamma_6^3\rho_6^{-1}.
\label{eq2}
\ee
Combining eqs.(\ref{eq1}) and (\ref{eq2}) we can estimate the NT gap hight as $h_3\sim 1.2\times10^2
\rho_6^2/\gamma_6^3$. On the other hand, using the expression for the accelerating potential drop and taking into
account the screening factor we can express $\gamma_6$ as (GMZ06): $\gamma_6\sim4.6\times b P^{-1} h_3^2 \eta$. Then,
using the NT condition we finally get the following expression for the characteristic Lorentz factors of particles
accelerated in PSG
\be
\gamma_6\sim 5 \left( P^{-1}b\eta\rho_6^{4}\right)^{\frac{1}{7}}.
\label{gamma}
\ee

\section{Emission and absorbtion of the curvature photos}

The power of the curvature radiation writes:
\be
L_{\rm cr}=4.6\times10^{-9}\frac{\gamma^4}{\rho^2}=4.6\times10^3 \frac{\gamma_6^4}{\rho_6^2}\;\;\;{\rm erg s^{-1}} .
\label{power}
\ee
Let us define $\bar L_{\rm cr}$  as:
\be
\bar L_{\rm cr}\equiv\frac{L_{\rm cr}}{\gamma m_e c^2}=5.6\times10^3 \frac{\gamma_6^3}{\rho_6^2}.
\label{power1}
\ee
Now we can claim that the particle can emit (by means of the curvature radiation) the energy, which is comparable with
its energy while passing the distance of the order of $\rho$  (i.e. during time $\rho/c$ ) if the condition
\be
\frac{\gamma_6^3}{\rho_6}\gtrsim 5.4
\label{cond1}
\ee
holds. It is clear from Fig.~\ref{curvature} and eq.~(\ref{gamma}) that the filed lines have enough curvature to
satisfy condition (\ref{cond1}). Fig.~\ref{curvature} shows that the particles accelerated in the outer ring of the
open field lines should pass the region where the curvature is high. The surface of this ring constitutes the major
part of the polar cap, therefore the particles created in the closed field line region carry energy comparable with the
energy of the outstreaming particles, i.e. comparable with the total energy available in IAR. At the same time the
energy of the curvature photons is about
\be
\hbar\omega_{\rm cr}\sim17\frac{\gamma_6^3}{\rho_6}m_e c^2,
\label{freq}
\ee
thus it is high enough for the pair creation.

Let us now estimate a free path of the curvature photons emitted at altitudes about few hundred meters from the stellar
surface. Using Erber's (1966) approximation and eqs. (\ref{cond1}) and (\ref{freq}) we can express the photon free path
as:
\be
l_{\rm ph}\sim4\times10^{-5}\left(\frac{B_{\rm q}}{B_{\perp}}\right)\exp\left(0.9\frac{B_{\rm q}}{B_{\perp}}\right)
\label{lphot}
\ee
Here $B_{\rm q}=4.414\times10^{13}$ G is so called quantum magnetic field and $B_{\perp}$ is a component of the
magnetic field which is perpendicular in respect to the photon propagation direction. The components of the magnetic
filed at the altitude where the curvature reaches its maximum are plotted in Fig.~\ref{magfieldup}. One can see that
the photons should be emitted almost along the radial component $B_r$ of the magnetic filed as $B_{\theta}$ is much
smaller. Therefore, the photon free path depends on $B_{\theta}\sim B_{\perp}$. Estimations show that $l_{\rm ph}$
varies from few hundred meters ($B_{\perp}/B_{\rm q}=0.05$) up to a few tens of kilometers.

\section{Discussion and Conclusion}

We have demonstrated that if the crust anchored magnetic anomalies at the surface of the neutron star are strong enough
to alter the curvature of the magnetic field lines, there exist favorable conditions for pair production in the closed
field line region. All those pairs move along the field lines, reach the stellar surface in the region of close field
lines and heat it. The hot surface emits the thermal radiation with the characteristic temperature that is defined by
the energy balance between the particle flux and the bolometric power. Therefore, the observed thermal radiation comes
not only from the polar cap region but also from the nearby area. For the parameters of PSR J1119 - 6127, the particle
flux can have enough power to support the observed thermal emission if eq.~(\ref{cond1}) is satisfied, which seems
quite reasonable.

The efficiency of the thermal X-ray emission is defined in the same way as it is presented by GMZ06, i.e. the
efficiency depends on the potential drop of the inner acceleration region. If the bolometric surface is smaller than
conventional polar cap area, then the characteristic temperature of the blackbody fit is also defined by the parameters
of the partially screened gap, i.e. by the thermostatic regulation of the screening parameter (Gil et al. 2003). In the
case of PSR J1119 - 6127 the blackbody temperature of the X-ray emission is simply defined from the heating-cooling
balance of the hot surface. The radiation comes from the closed field line region, where the electric field is always
perpendicular to the magnetic field, consequently there no partial screening is possible. We believe that it is another
argument in favour of the partially screened gap model for the inner acceleration region of pulsars.

\begin{acknowledgements}
We gratefully acknowledge the support by the WE-Heraeus
foundation. We also acknowledge the support of the Polish State
Committee for scientific research under Grant 2 P03D 029 26.
\end{acknowledgements}

{}
      \clearpage

\end{document}